\pdfoutput=1

\documentclass{sig-alternate-05-2015-arxiv}
\usepackage{txfonts}
\usepackage{color}
\definecolor{maroon}{rgb}{.66,.30,.27}
\definecolor{cyan}{rgb}{.39,.79,.87} 
\usepackage{xr-hyper}
\usepackage[singlelinecheck=false]{subcaption}
\usepackage{mathtools}
\usepackage[table,xcdraw]{xcolor}
\usepackage{arydshln}
\usepackage{tabu}
\usepackage{array,multirow,graphicx}
\usepackage{booktabs}
\usepackage{sfmath}
\usepackage{tabularx}
\usepackage{cleveref}
\usepackage[colorlinks]{hyperref}
\usepackage[labelfont=bf, textfont=bf]{caption}
\usepackage{balance}
\usepackage{colortbl}
\usepackage{hhline}
\usepackage{enumitem}
\usepackage{boldline}

\begin{document}






%

\title{Credibility and Dynamics of Collective Attention}
\def\sharedaffiliation{%
\end{tabular}
\begin{tabular}{c}}
\numberofauthors{3}
    \author{
      \alignauthor Tanushree Mitra\\      
      \email{tmitra3@gatech.edu}
      \alignauthor Graham Wright\\     
      \email{gwrong@gatech.edu}
      \alignauthor Eric Gilbert\\    
      \email{gilbert@cc.gatech.edu}
      \sharedaffiliation
      \affaddr{School of Interactive Computing \& GVU Center}  \\
      \affaddr{Georgia Institute of Technology} \\
      \affaddr{Atlanta, GA, USA}
          }
%


\maketitle
\begin{abstract}
Today, social media provide the means by which billions of people experience news and events happening around the world. However, the absence of traditional journalistic gatekeeping allows information to flow unencumbered through these platforms, often raising questions of veracity and credibility of the reported information. Here we ask: How do the dynamics of collective attention directed toward an event reported on social media vary with its perceived credibility? By examining the first large-scale, systematically tracked credibility database of public Twitter messages (47M messages corresponding to 1,138 real-world events over a span of three months), we established a relationship between the temporal dynamics of events reported on social media and their associated level of credibility judgments. Representing collective attention by the aggregate temporal signatures of an event reportage, we found that the amount of continued attention focused on an event provides information about its associated levels of perceived credibility. Events exhibiting sustained, intermittent bursts of attention were found to be associated with lower levels of perceived credibility. In other words, as more people showed interest during moments of transient collective attention, the associated uncertainty surrounding these events also increased. 
\end{abstract}

%
%

%
%

%
%



\section{Introduction}
Online social networks act as information conduits for real-world news and events \cite{pew}, largely driven by collective attention from multiple social media actors \cite{wu2007novelty}. 
Collective human attention drives various social, economic and technological phenomenon, such as herding behavior in financial markets \cite{sinha2010econophysics}, formation of trends \cite{asur2011trends}, popularity of news \cite{wu2007novelty}, web pages \cite{ratkiewicz2010characterizing}, 
and music \cite{salganik2006experimental}, propagation of memes \cite{leskovec2007dynamics}, ideas, opinions and topics \cite{romero2011differences}, person-to-person word-of-mouth advertising and viral marketing \cite{leskovec2007dynamics}, and diffusion of product and innovation \cite{bass1969frank}. Moreover, it is the key phenomenon underlying social media reporting of emerging topics and breaking news \cite{mathioudakis2010twittermonitor}. 
However, unlike traditional news media---where information is curated by experienced journalists---social media news is unfiltered and therefore not subject to the same verification process as information presented by way of conventional sources. This naturally calls into question its credibility and the means with which to assess its credibility. Although scholars have increasingly expressed concern over the threats posed by digital misinformation, ranging from panic and violence incitement in society to libel and defamation of individuals or organizations \cite{howell2013digital}, questions concerning the relationship between collective attention and information credibility have not been systematically quantified.

A fundamental attribute underlying any collective human behavior is how that behavior unfolds over time \cite{barabasi2005origin, crane2008robust}. 
Is there a relationship between allocation of collective attention and perceived credibility of events reported through social media? Do occasional bursts in collective attention---as more eyes and voices are drawn to the event's reportage---correspond to less certain information concerning the event? Uncovering the relationship between collective human behavior and information credibility is important for assessing the veracity of event reportage as it unfolds on social media. This relationship, if it exists, can provide insights into ways to disambiguate misinformation from accurate news stories in social networks---a medium central to the way we consume information \cite{pew} and one where digital misinformation is pervasive \cite{del2016spreading}. 

Empirical attempts at answering these questions in naturalistic settings have been constrained by difficulties in tracking social media posts in conjunction with judgments concerning the accuracy of the underlying information. Previous studies have instead focused on individual case studies involving specific news events \cite{maddock2015characterizing, liu2014rumors, sydney2016}, or have retrospectively studied a set of multiple prominent events \cite{ladaicwsm, del2016spreading} which were known to contain misinformation. While useful, these approaches raise sampling concerns. In particular, they are based on the post-hoc investigation of events with known credibility levels, and thus select on the dependent variable \cite{tufekci2014big}.
Although these studies suggest the possibility of spikes in collective attention when false rumors propagate through social networks, the relation between collective attention and information credibility has not been systematically tested. 

We tested this relation by analyzing data from the first large-scale, longitudinal credibility corpus, called CREDBANK \cite{mitra2015credbank}. The massive dataset was constructed by iteratively tracking millions of public Twitter posts. 
Twitter is a microblogging site where people write short time-stamped messages publishing their daily social activities \cite{naaman2010really} or discussing world events \cite{kwak2010twitter}. 
Hence, tweeting activity comprises regular circadian rhythms \cite{golder2011diurnal} intertwined with irregular bursts of activities corresponding to real-world news events \cite{mathioudakis2010twittermonitor}. In recent years, Twitter has become an attractive source for disseminating information pertaining to news events \cite{kwak2010twitter}. CREDBANK's credibility corpus is based on tracking all large-scale, real-world events surfacing on Twitter between October 2014 and February 2015, followed by credibility assessment through a verified human annotation framework. This iterative framework resulted in an experimental setup that captured accurate human judgments of credibility of social media information soon after it gained collective attention. It is important to note that while this process cannot arrive at the \emph{truth} of the event reporting (perhaps an impossible epistemological task), it does capture expert-level human judgment at the time the event unfolded.
CREDBANK contains 1,377 social media events collected over a period of three months, along with 66 million tweets nested within the event topics. The uniqueness of the dataset is evident not only from the systematic collection process but also from the range of the collected events.  It contains, for example, objections to red cards thrown soccer matches, as well as the emergence of Ebola in West Africa. 

Although the nature of this data limits causal inference, we were able to test the correspondence between collective attention and the level of information credibility. 
After filtering out unique event instances, we were left with a pruned corpus of 1,138 real-world events spread over 47M tweets. Analyzing this massive dataset, we find that the amount of recurring collective attention bursts could be used to determine the level of perceived credibility of an event. Specifically, we demonstrate that multiple occasional bursts of collective attention toward an event is associated with lower levels of perceived credibility. This finding opens a new perspective in the understanding of human collective attention and its relation to the certainty of information. In doing so, our results can have widespread implications in fields where predictive inference based on online collective interests dictates economic decisions, emergency responses, resource allocation or product recommendations \cite{goel2010predicting, wu2010evidence}; hence trusting the credibility of the collective reports is essential for an accurate anticipation by the predictive process. 

\section{Related Work}
\subsection{Social Media and Credibility}
With social media's rise as an important news source \cite{pew}, individuals are constantly relying on online social networks to consume and share information, without recourse to official sources. 
However, modern online social networks like Facebook and Twitter are neutral with respect to the quality of information \cite{ladaicwsm}. Moreover, users of these sites have been found to be poor judges of information credibility based on content alone \cite{morris2012tweeting}. 
Thus, scholars have increasingly become interested in assessing the credibility of social media content.
Studies have focused on investigating specific events that were subjects of misinformation, such as the spread of rumors during the 2011 Great East Japan earthquake \cite{liu2014rumors}, the 2013 Boston marathon bombings \cite{maddock2015characterizing} and the 2014 Sydney siege event \cite{sydney2016}. 
Studies have also engaged in extensive analysis of multiple historically reported cases of rumor, such as, automatically classifying rumor instances  \cite{zeng2016unconfirmed} or predicting the credibility level of tweets \cite{castillo2011information, gupta2014tweetcred, qazvinian2011rumor}.
However, these studies are based on the retrospective investigation of popular historical events. Hence, they suffer from the \emph{selection on the dependent variable} confound. On the contrary, our study overcomes this confound by grounding its results on CREDBANK's data \cite{mitra2015credbank} -- a credibility corpus which asks human raters to assess the credibility of \emph{all} social media events in near real-time.

\subsection{Collective Attention}
A phenomenon which is vital towards the spread of social media information is ``collective attention"  \cite{ratkiewicz2010characterizing}. 
Hence, researchers have been attracted toward understanding how attention to new information propagates among large groups of people. 
While some studies have shown that dynamics of collective attention of online content is characterized by bursts signifying popularity changes \cite{lehmann2012dynamical, ratkiewicz2010characterizing}, others have demonstrated a natural time scale over which attention fades \cite{wu2007novelty}. A study investigating the emergence of collective attention on Twitter, found that although people's attention is dispersed over a wide variety of concerns, it can concentrate on particular events and shift elsewhere either very rapidly or gradually \cite{sasahara2013quantifying}.
Another parallel study focusing on spikes of collective attention in Twitter, analyzed the popularity peaks of Twitter hashtags \cite{lehmann2012dynamical}. They found that the evolution of hashtag popularity over time defined discrete classes of hashtags. Drawing on the progress of these studies, we ask:  does the process of evolving collective attention reflect the underlying credibility of a social media story? Unraveling the relation between collective attention rhythms and corresponding credibility level is a complex empirical problem. It requires longitudinal tracking of collective mentions of newsworthy stories in social media along with their in-situ credibility judgments. To that end, CREDBANK provides the most consistently tracked social media information and its associated credibility scores.

\subsection{Time Matters}

One useful way to understand the interplay between collective attention and information credibility is to examine user activity and information patterns through the lens of time. For decades social scientists have investigated the timing of individual activity to understand the complexity of collective human action. They have reported that timing can range from random \cite{haight1967handbook} to well correlated bursty activity patterns \cite{barabasi2005origin}. The bursts in human collective action have not only led to social media reporting of emerging topics, but have also exhibited rich temporal dynamics of social media information spread \cite{mathioudakis2010twittermonitor}. For example, information diffusing through micro-blogging platforms like Twitter have  demonstrated a short life span \cite{wu2007novelty}, with content rising and falling in popularity within hours; whereas, short quoted phrases (known as \emph{memes}) have displayed several days to rise and fade away \cite{leskovec2009meme}. On the other hand, general themes (like `politics', `economy', `terrorism') have shown an even larger temporal life span \cite{gruhl2004information, wang2007mining}. Social psychologists studying the spread of news and rumor have also noted the importance of temporal patterns in rumor transmission -- different types of rumor mongering statements persist over varying temporal spans \cite{bordia1998rumor, shibutani1966improvised}. However, despite the importance of temporal patterns in information diffusion and rumor transmission, there has been little work in understanding temporal trends in events and its associated credibility assessments. This paper is a step towards unraveling that relation. 

\section{Method}
\subsection{Data Description}
The data investigated in this work was gathered from the CREDBANK corpus \cite{mitra2015credbank} which we had built to systematically study social media credibility. The corpus contains 1,377 events as they surfaced on Twitter between October 2014 and February 2015, their corresponding public tweets (a total of 66M messages) and their associated credibility ratings.
We built CREDBANK by iteratively tracking millions of public posts streaming through Twitter; computationally detecting the underlying topics (i.e. clusters) of discussion in every block of million tweets; separating event-specific topics from non-event topics by asking independent human raters from Amazon Mechanical Turk (AMT); and then for each of these event topics we gathered credibility ratings on a 5-point Likert scale ranging from `Certainly Inaccurate' (-2) to `Certainly Accurate' (+2). Each event is represented by a combination of top three terms from that event topic. For example, ``chelsea", ``game", ``goal" refers to a football match event at a point in time when Chelsea scored a goal. 
Thirty independent human raters from AMT judged the accuracy level of an event by browsing through tweets in real-time, where each tweet contained all the top three topical terms. Such a task design ensured that the annotation task closely mimics the way a person would search Twitter to see information related to an event. Moreover, limiting tweets containing all top 3 topical terms ensured a balance between being too generic (by including fewer terms) and too specific (by including more terms), and also provided enough context to a human rater for performing the task.

To guarantee that our collected ratings is at par with expert level judgements, we performed multiple controlled experiments before finalizing the strategy best suited for obtaining quality annotations \cite{mitra2015comparing}. Finally, we searched Twitter to collect all tweets specific to the event topic from the past seven days (a time limit enforced by Twitter's search program interface). This did not seem to be a limitation because our experimental setup was tracking recent events. Moreover, research has demonstrated that news worthy topics on twitter have an active period of at most a week \cite{kwak2010twitter}.  Overall, this iterative framework resulted in a natural experimental setup where the credibility of social media information was being tracked soon after it gained collective attention. Additional details of our data collection process is outlined in the Appendix. A representative sample of events tracked during this three month period along with their credibility ratings is presented in Table \ref{tab:event}. By listing the range of diverse events, our aim is to demonstrate the richness of the dataset and hence the generalizability of our results ensuing from this dataset. For example, it contains events ranging from celebrity deaths and terroristic attacks to missing airplanes and soccer matches.
\hypersetup{
    urlcolor=blue           
}
\renewcommand{\thefootnote}{*}
    \footnotetext{\noindent  \textbf{NOTE}: Credibility assessment is a long running project by the authors of this submission. It started with the creation of CREDBANK corpus \cite{mitra2015credbank} and has since been followed with multiple analysis of this corpus \cite{mitra2017credlang}. For the ease of the reader, materials and methods have been reproduced in its abbreviated form in the current submission. The analysis, results and the overall contributions of this paper are new.}

\begin{table}[!t]
\begin{tabular}{llrr}
\toprule
\textbf{Cred Class} 	& \textbf{$P_{ca}$ Range}	& \textbf{Total Events} 	& \textbf{Distinct Events} 	\\ \midrule
Perfect    		& $0.9 \leq P_{ca} \leq 1.0$     	& 421           			& 342          		 	\\
High       		& $0.8 \leq P_{ca} < 0.9$ 		& 433          			& 337          		 	\\
Moderate   	& $0.6 \leq P_{ca} < 0.8$ 		& 414           			& 358          		 	\\ 
Low        		& $0.0 \leq P_{ca} < 0.6$    	& 109           			& 101   			 	\\
\bottomrule
\end{tabular}
\caption{Credibility classes and corresponding event counts. ``Total Events'' column shows event counts from CREDBANK. ``Distinct Events'' column lists counts from the pruned corpus.}
\label{tab:credclass}
\vspace{-15pt}
\end{table}

\subsection{Pruning Corpus for Sample Independence}
During the iterative building of CREDBANK, if an event trended on Twitter for a sufficiently long time period, it is possible that the event is curated multiple times. For example, the event "arsenal", "win", "city" corresponds to the Arsenals winning the football match against Stoke city. People on Twitter had active conversations about the event for several hours, resulting in the event being captured more than once in CREDBANK. However, our statistical analysis (discussed shortly) required sample independence. Occurrence of multiple instances of the same event will likely violate the independence assumption. Hence, we pruned our event sample to keep single distinct instances of each event. By matching the three terms in each event topic, we looked for duplicate event occurrences. Thereafter, if multiple instances of the same event existed, we picked the event which had the earliest curation time. Restricting events by earliest curation times ensured that we retained crowd worker annotations corresponding to the very first time that they performed the annotation task; hence preventing any potential prior knowledge bias. Our pruning step resulted in a dataset of 1,138 events spanning 47,000,127 tweets.

\subsection{Credibility Classification}
We measured an event's perceived credibility level based on how many human raters agreed that the event was ``Certainly Accurate''. More formally, for each event we find the proportion $P_{ca}$ of ratings marked as ``Certainly Accurate''.
\begin{equation*}
P_{ca} = \frac{\textit{Count ``Certainly Accurate'' ratings for an event}}{\textit{Total ratings for that event}}
\end{equation*}

To have a reasonable comparison it is impractical to treat $P_{ca}$ as a continuous variable and have a category corresponding to every value of $P_{ca}$. Hence, we placed $P_{ca}$ into four classes that cover a range of values (see Table \ref{tab:credclass}). The class names are based on the perceived degree of accuracy of the event in that class. For example, events belonging to the ``Perfect Credibility'' class were rated as ``Certainly Accurate'' by almost all raters ($0.9 \leq P_{ca} < 1$). 


\begin{figure}[!t]
\centering
\includegraphics[width=\linewidth]{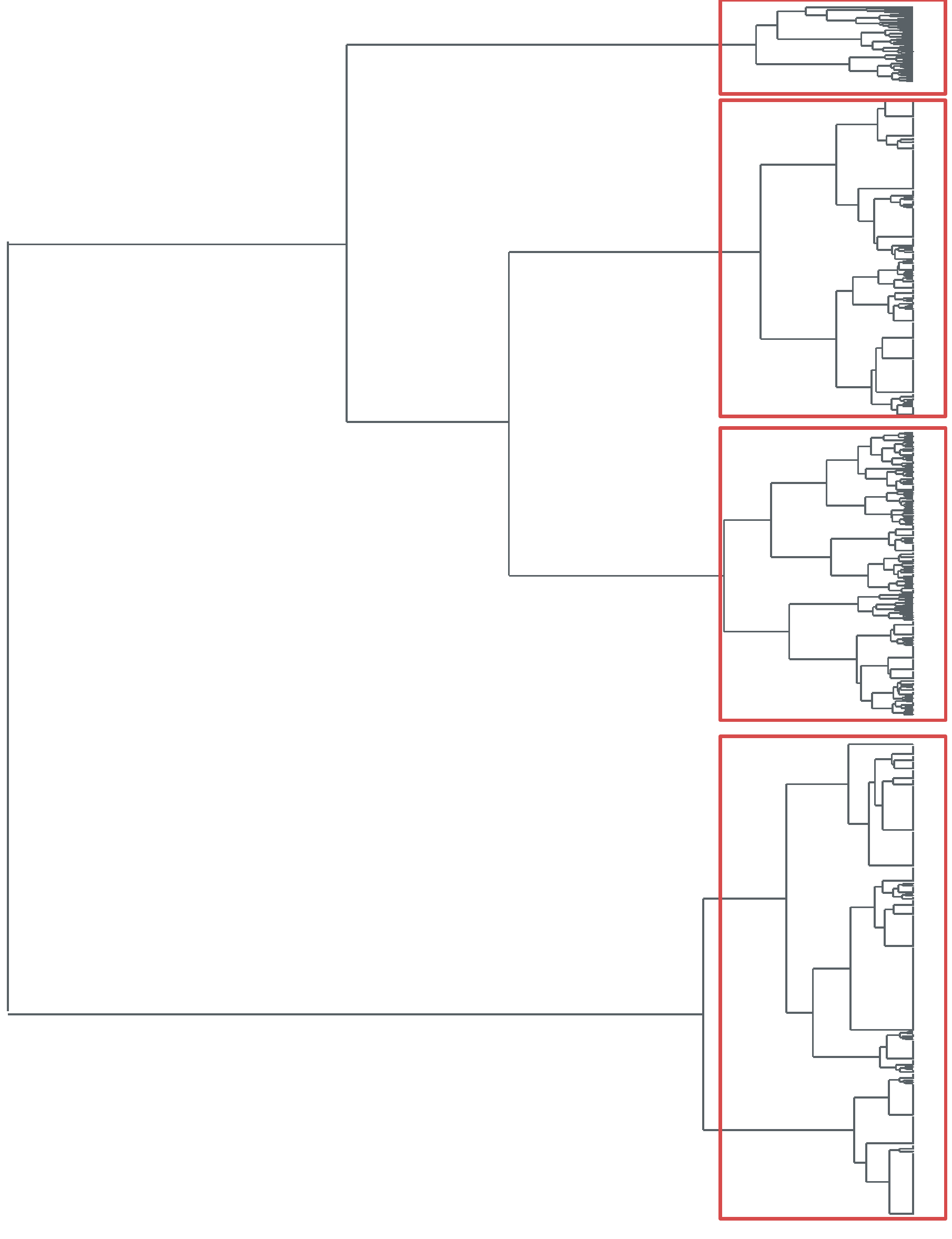}
\vspace{-20pt}
\caption{Dendogram from hierarchical clustering of the events from CREDBANK. The boxes show the four clusters.}
\vspace{-15pt}
\label{fig:hclust}
\end{figure}

\begin{table*}[t!]
\vspace{-0.5cm}
\captionsetup{width=.98\textwidth}

\begin{center}
{ \sffamily \small
\begin{tabularx}{.98\textwidth}{Xrllrr}
\hlineB{3}
\textbf{Event Terms}               & \textbf{\# Tweets} & \textbf{Start time} & \textbf{End Time}      & \textbf{Ratings}   & \textbf{$P_{ca}$} \\ \hlineB{3}
\multicolumn{6}{c}{\cellcolor[HTML]{EFEFEF}\textbf{Perfect Credibility: \boldmath$0.9 \leq P_{ca} \leq 1$}}                                                                                            \\ \hline
george clooney \#goldenglobes      & 10350                & 2015-01-12 08:50    & 2015-01-12 18:10       & {[}0 0 1 1 28{]}   & 0.93           \\
king mlk martin                    & 88045                & 2015-01-15 22:00    & 2015-01-15 22:00       & {[}0 0 0 2 28{]}   & 0.93           \\
win pakistan test                  & 5478                 & 2014-10-26 18:10    & 2014-11-03 21:00       & {[}0 0 0 3 27{]}   & 0.90            \\
george arrested zimmerman                   & 45645                & 2015-01-07 19:40    & 2015-01-11 00:50       & {[}0 0 0 3 27{]}   & 0.90   \\
scott rip sad                          & 26006 & 2014-12-29 07:50 & 2015-01-05 18:10 & {[}0 0 0 3 27{]} & 0.90 \\ 
\hline
\multicolumn{6}{c}{\cellcolor[HTML]{EFEFEF}\textbf{High Credibility: \boldmath$0.8 \leq P_{ca} < 0.9$}}                                                                                               \\ \hline
beckham odell catches              & 21848                & 2014-11-04 04:10    	& 2014-11-04 22:20       	& {[}0 0 0 4 26{]}   & 0.87           \\
eric garner death                	& 180582 		    & 2014-11-26 08:30 	& 2014-12-04 07:10 		& {[}1 1 0 2 26{]}  & 0.87  \\
windows microsoft holographic    & 18306 & 2015-01-21 23:40 & 2015-01-25 10:00 & {[}0 0 0 4 26{]}  & 0.87  \\
kayla mueller isis               & 65819 & 2015-02-06 21:10 & 2015-02-12 00:10 & {[}0 0 0 8 52{]}  & 0.87 \\
liverpool arsenal goal             & 16713                & 2014-12-14 05:20    & 2014-12-14 05:20    & {[}0 1 0 4 25{]}   & 0.83           \\
\hline
\multicolumn{6}{c}{\cellcolor[HTML]{EFEFEF}\textbf{Moderate Credibility: \boldmath$0.6 \leq P_{ca} < 0.8$}}                                                                                           \\ \hline
children pakistan \#peshawarattack & 24239                & 2014-12-16 12:30    & 2014-12-17 20:10       & {[}0 1 1 5 23{]}   & 0.77           \\
\#ericgarner protesters police     & 12510                & 2014-12-04 00:50    & 2014-12-05 10:20       & {[}0 0 2 6 22{]}   & 0.73           \\
sydney hostage \#sydneysiege       & 21835                & 2014-12-15 04:20    & 2014-12-15 17:20       & {[}0 0 2 6 22{]}   & 0.73           \\
bobby shmurda bail                 & 22362 & 2014-12-17 21:40 & 2014-12-19 17:30 & {[}0 0 1 7 22{]}   & 0.73  \\
\#antoniomartin ambulance shot             &  6330 & 2014-12-24 11:30 & 2014-12-24 23:10 & {[}0 0 3 9 18{]}    & 0.60          \\ 
\hline
\multicolumn{6}{c}{\cellcolor[HTML]{EFEFEF}\textbf{Low Credibility: \boldmath$0 \leq P_{ca} < 0.6$}}  \\ \hline
gerrard liverpool steven               &  204026  & 2014-12-26 03:40 & 2015-01-02 20:20 & {[}0 1 3 9 17{]}    & 0.57  \\
\#chapelhillshooting muslim white     & 35282       & 2015-02-11 11:20 & 2015-02-13 06:20 & {[}2 2 8 16 32{]}   & 0.53  \\

paris boko killed                  & 3917                 & 2015-01-07 22:50    & 2015-01-11 01:50       & {[}0 3 1 11 15{]}  & 0.50            \\
ebola \#ebola travel               & 27796                & 2014-10-09 06:10    & 2014-10-17 09:10       & {[}2 2 6 10 10{]}  & 0.33           \\
killed hostage isis                & 25925                & 2015-01-31 20:20    & 2015-02-08 10:00       & {[}0 8 22 14 16{]} & 0.27           \\
\hlineB{3}
\end{tabularx}}
\end{center}
\vspace{-10pt}
\caption{Sample events from the CREDANK corpus grouped by their credibility classes. Events are represented with three event terms. Start and end times denote the time period during which tweets were collected using Twitter's search API combined with a search query containing a boolean \emph{AND} of all three event terms. Rating shows the count of Turkers who selected an option from the 5-point Likert scale ranging from -2 (``Certainly Inaccurate'') to +2 (``Certainly Accurate''). 
} 
\label{tab:event}
\vspace{-0.3cm}
\end{table*}

\subsection{Validating credibility classification}
To ensure that our $P_{ca}$ based credibility classification is a reasonable classification, we compared classes generated by our $P_{ca}$ method against those obtained via data-driven classification.

\vspace{25pt}
\subsubsection{Generating data-driven credibility classes} 
We used hierarchical agglomerative clustering (HAC) \cite{maimon2005data} to generate data-driven classes of the credibility rating distributions.
HAC is a bottom-up clustering approach which starts with each observation in its own cluster followed by merging pairs of clusters based on a similarity metric. In the absence of a prior hypothesis regarding the number of clusters, HAC is the preferred clustering method. HAC-based clustering approach groups the events based on the shape of their credibility curves on the 5-point Likert scale. Such shape based clustering approach has been used in prior work to cluster based on the shape of popularity peaks \cite{crane2008robust, yang2011patterns}.
We used the Euclidean distance similarity metric and Ward's fusion strategy for merging \cite{ward1963hierarchical}. The choice of this strategy minimizes the within-cluster variance thus maximizing within-group similarity \cite{ward1963hierarchical}. Figure \ref{fig:hclust} shows the resulting dendogram from hierarchical clustering. The boxes correspond to the credibility groups when the dendogram is cut into four clusters.

\vspace{-5pt}
\subsubsection{Comparing $P_{ca}$ classes to HAC-based classes}
Is the $P_{ca}$ based credibility classification a close approximation of the HAC based classification? Essentially, we need a metric to compare two clusterings of the same dataset. In other words, we need to measure how often both clustering methods classify the same set of observations as members of the same cluster. We borrow a technique proposed by Tibshirani et al. \cite{tibshirani2005cluster}. 
Let $P_{clust} = \{{x_1}_{c_1}, {x_2}_{c_1}, {x_3}_{c_2}, \cdots , {x_n}_{c_4}\}$ denote the cluster labels from $P_{ca}$ based classification and $H_{clust} = \{{x_1}_{h_1}, {x_2}_{h_3}, {x_3}_{h_3}, \cdots , {x_n}_{h_4}\}$ the labels from HAC-based classification of the same dataset $D$ of $n$ observations. Here, ${x_i}_{c_j}$ denotes that the $i^{th}$ observation belongs to cluster $c_j$ as per the $P_{ca}$ classification and ${x_i}_{h_j}$ denotes that the $i^{th}$ observation belongs to cluster $h_j$ as per the HAC classification. We see that ${x_1}_{c_1}$ and ${x_2}_{c_1}$ belong to the same cluster. Such pairs are called ``co-members''. While (${x_1}_{c_1}$, ${x_2}_{c_1}$)  are co-members as per $P_{ca}$ classification,  (${x_2}_{h_3}, {x_3}_{h_3}$) are co-members from HAC classification. For each clustering method, we first compute all pairwise co-memberships. Next, we measure agreement between the clustering methods by computing the Rand similarity coefficient (R) from co-memberships.
\begin{equation*}
R = \frac{N_{11} + N_{00}}{N_{11} + N_{10}+ N_{01}+ N_{00}}
\end{equation*}
\begin{description}
\setlength{\itemsep}{0cm}%
  \setlength{\parskip}{0cm}%
  \item[$N_{11}$]: number of observation pairs where both are co-members in both clustering methods.
  \item[$N_{10}$]: number of observation pairs where the observations are co-members in the first clustering
method, but not in the second.
  \item[$N_{01}$]: number of observation pairs where the observations are co-members in the second clustering
method, but not in the first.
  \item[$N_{00}$]: number of observation pairs where neither pair is co-member in either clustering method.
\end{description}

\begin{figure*}[t!]
\centering
 \includegraphics[width=0.90\linewidth]{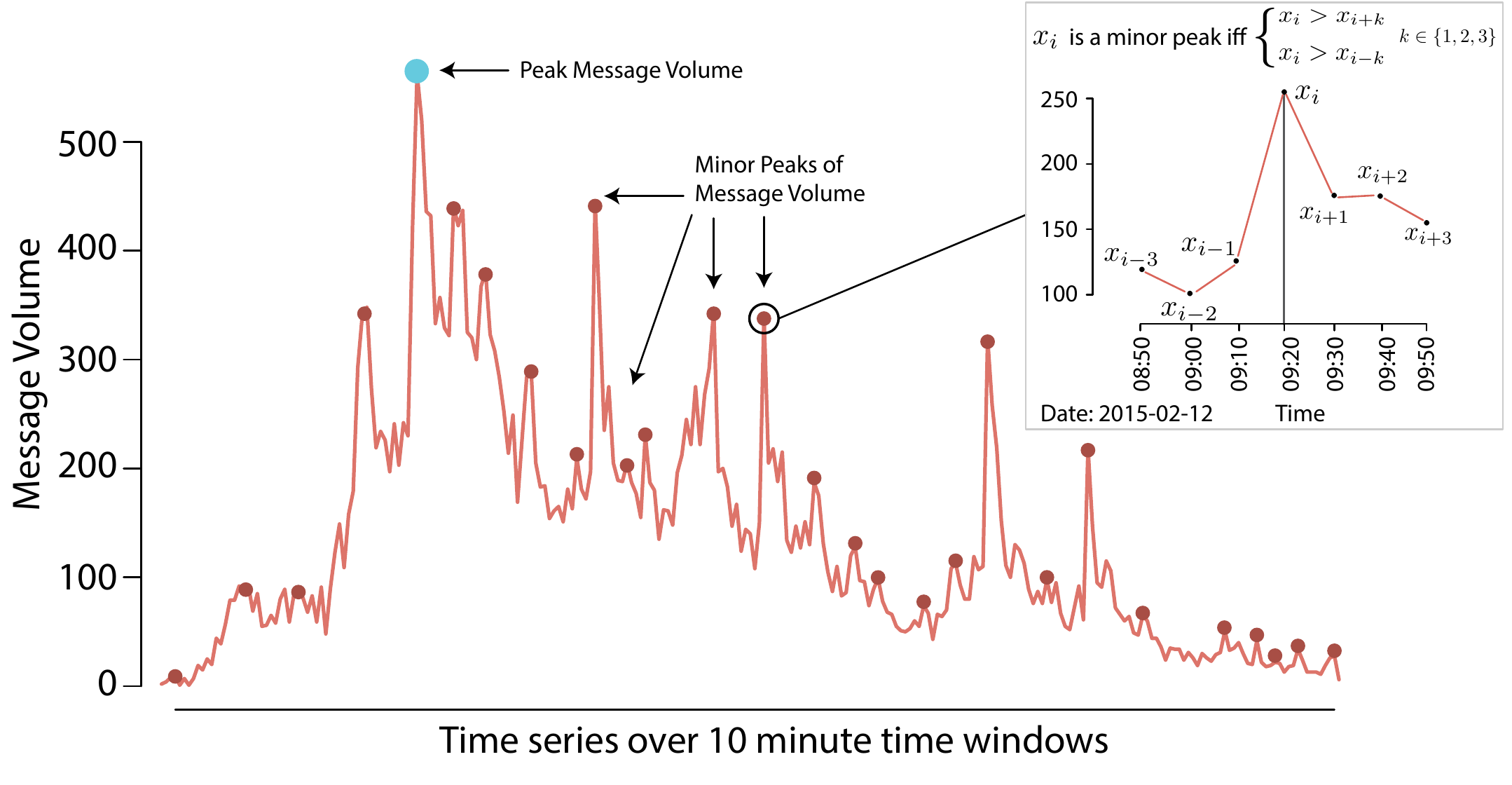}
\vspace{-10pt}
\caption{The time series of message volume for a sample event reported on Twitter. The event corresponds to Twitter discussions, where each tweet contained all three terms: ``\#chapellhillshooting'', ``muslim'' and ``white''. The \textcolor{cyan}{$\medbullet$} dot corresponds to the time window having maximum \emph{message volume} while the \textcolor{maroon}{$\medbullet$} dots correspond to the minor peaks observed in this volume. The inset diagram on the right side zooms in on one of the minor peaks, along with the rule triggering its designation.} 
\label{fig:peaklabels}
\vspace{-10pt}
\end{figure*}

Rand similarity coefficients range between 0 and 1, with 1 corresponding to perfect agreement between the two clustering methods. We obtain a fairly high R of 0.774 denoting high agreement between our $P_{ca}$ based and HAC-based clustering approaches. We favor our proportion-based ($P_{ca}$) clustering technique over data-driven approaches because the former is much more interpretable and readily generalizable and adaptable to domains other than Twitter on which CREDBANK was constructed.

\section{Statistical Measures}
To understand the relation between collective attention and information credibility, we computed our measures using time-stamped tweets from the CREDBANK corpus, where groups of tweets corresponded to discussions of an event by multiple Twitter users over a certain time span. 
\subsection{Collective Attention Metrics}
Collective attention of such event reportage was measured using two metrics: 
\begin{enumerate}[noitemsep,nolistsep, topsep=0pt, parsep=0pt,partopsep=0pt]
\itemsep0em 
\item \emph{Message Volume}: message volume tracks the aggregate number of messages over time
\item \emph{People Volume}: people volume records the aggregate count of unique users paying attention to the story over time. 
\end{enumerate}
\vspace{5pt}
Each measure is represented as a time series with message (or unique user) counts aggregated over 10-minute time intervals. Our choice of a 10-minute window is supported by studies showing that Twitter acts as a medium for reporting breaking news and hence is characterized by fast diffusion of information \cite{hu2012breaking, kwak2010twitter}. Thus, tracking collective attention on the order of minutes is a reasonable representation of a rapidly evolving phenomenon. Each event may differ in the temporal dynamics of its collective attention; thus inferences drawn on a small set of events tracked for a few days may be confounded by temporal traits peculiar to certain news stories. However, by tracking news stories over several months and averaging over hundreds of such collective attention rhythms, our results represent the most consistent relations between the dynamics of collective attention and perceptions of information credibility.
Our rationale for using \emph{people volume}, in addition to \emph{message volume}, as a collective attention metric is to ensure that the collective attention measured is not confounded by superfluous posting activity from potential Twitter bots, automated programs posing as human beings \cite{chu2010tweeting}. Since \emph{people volume} corresponds to the unique number of individuals paying attention to the event over time, it aims to counteract any extreme posting activity by such bots. 

As an example illustrating our data and methods, Figure \ref{fig:peaklabels} shows aggregate message volume for an event reported on Twitter where every message contained the terms ``\#chapellhillshooting'', ``muslim'' and ``white''. On February 10, 2015, three Muslim students in Chapel Hill, North Carolina were shot to death by a white neighbor  and speculations concerning the motives of the shooter surrounded the event \cite{time_chappell}. While authorities suggested the motive to be an ongoing dispute between neighbors over a parking space, many social media users suggested a hate crime as the motive. Twitter messages concerning this specific topic on February 12 blamed the media for ignoring the coverage of an event involving Muslim killings and suggested the shooting was an act of terrorism and so a hate crime. It was not until February 13 that authorities opened an investigation to determine if the shooting was in fact a hate crime. Credibility rating distributions showed that less than 50\% of raters agreed that the social media reportage of the event was ``Certainly Accurate'', thereby questioning the alleged terror claims underlying the act. 

\subsection{Temporal Measures of Collective Attention}
To quantify the importance of the time when collective attention maximized, we first computed the strict global maximum in the time series \cite{stewart2015calculus}. We call this the \emph{peak attention}. This is the ratio of messages (or unique people) within the peak time window to the total cumulative volume of messages (or unique people) over the entire event time series:

\vspace{-10pt}
\begin{align}
\textit{Peak Attention} = \frac{max(x_1, \cdots, x_n)}{\sum\limits_{i=1}\limits^{n}x_i}
\vspace{-10pt}
\end{align}
where $x_i$ is the count of messages (or unique people) in time window $i$ in an event time series $x_1, x_2, \cdots, x_n$.
Our choice of the above measure is based on the success of prior studies using peak fraction based metrics to successfully characterize herding behavior over time \cite{crane2008robust, yang2011patterns}. 
To illustrate how peak attention measure can characterize variations in time series, consider the example of an event reportage marked by a sudden spike in collective attention followed by a subsequent drop. The lack of precursory growth suggests that most of the attention was concentrated on the peak, thereby resulting in high \emph{peak attention} (Figure \ref{fig:timeseries}c and  \ref{fig:timeseries}d). Whereas, an event with steady growth in collective attention, followed by a gradual decay would imply a relatively smaller fraction of attention in the peak, thus leading to lower value of \emph{peak attenion} (Figure \ref{fig:timeseries}b).

While \emph{peak attention} captures the importance of the maximal burst in collective attention, it does not take into account the presence or absence of spikes in the precursory growth and in the subsequent decay following the burst. Hence, we define a measure to quantify the spikiness in collective attention. We detect all strict local maxima \cite{stewart2015calculus} in each of the event time series. 
A strict local maxima corresponds to an instance in the time series when the volume of messages (or unique people) is larger than the volume in the neighboring time windows. We define this neighborhood as three time windows on either side of the local maxima and call these local maxima \emph{minor peaks}. Thus, the attention in a minor peak is higher than the attention 30 minutes (i.e., three time windows times 10-minute window size) before and after the occurrence of a peak.
\vspace{-5pt}
\begin{align}
x_i \textit{ is a minor peak iff } \left\{\begin{matrix}
x_i > x_{i+k}\\ 
x_i > x_{i-k}
\end{matrix}\right.,  k \in \left \{1, 2, 3 \right \}
\end{align}
We then define minor peak attention as the ratio of messages (or unique people) in the local maxima relative to the total cumulative volume of messages (or unique people) over the entire event time series. Formally, if $\mathcal{M}$ is the set of all minor peak indices in an event's message (or unique people) time series, then minor peak attention is defined as follows:
\vspace{-10pt}
\begin{align}
\textit{Minor Peak Attention} = \frac{\sum\limits_{j \in \mathcal{M}} x_j}{\sum\limits_{i=1}\limits^{n}x_i}
\vspace{-10pt}
\end{align}
The inset diagram in Figure \ref{fig:peaklabels} shows a local maximum. While the peak attention captures the maximum momentary interest that an event acquires during its lifetime on Twitter, the points representing minor peak attention reflect renewed and ongoing recurrences of momentary interest. 
Additionally, both these measures have two important properties: both are invariant with respect to scaling and shifting \cite{yang2011patterns}. First, since both measures are proportions based on cumulative collective attention, they are invariant to the overall volume of attention. Hence, two event time series having similar peaky shapes but different total attention volumes would be treated similarly. Secondly, both measures are computed independent of the maxima position on the time axis. Thus, if two event time series peaks occur at different times but possess a similar peaky structure, the measures will be invariant to the translations on the time axis. 
Hence, both these measures---despite being simple representations of temporal dynamics---are useful in interpreting the relationship between collective attention rhythms and event credibility across a range of different events exhibiting high variability in overall popularity and time of popularity.

\begin{figure*}[!t]
\centering
\begin{subfigure}[b]{0.49\textwidth}
 \includegraphics[width=\linewidth]{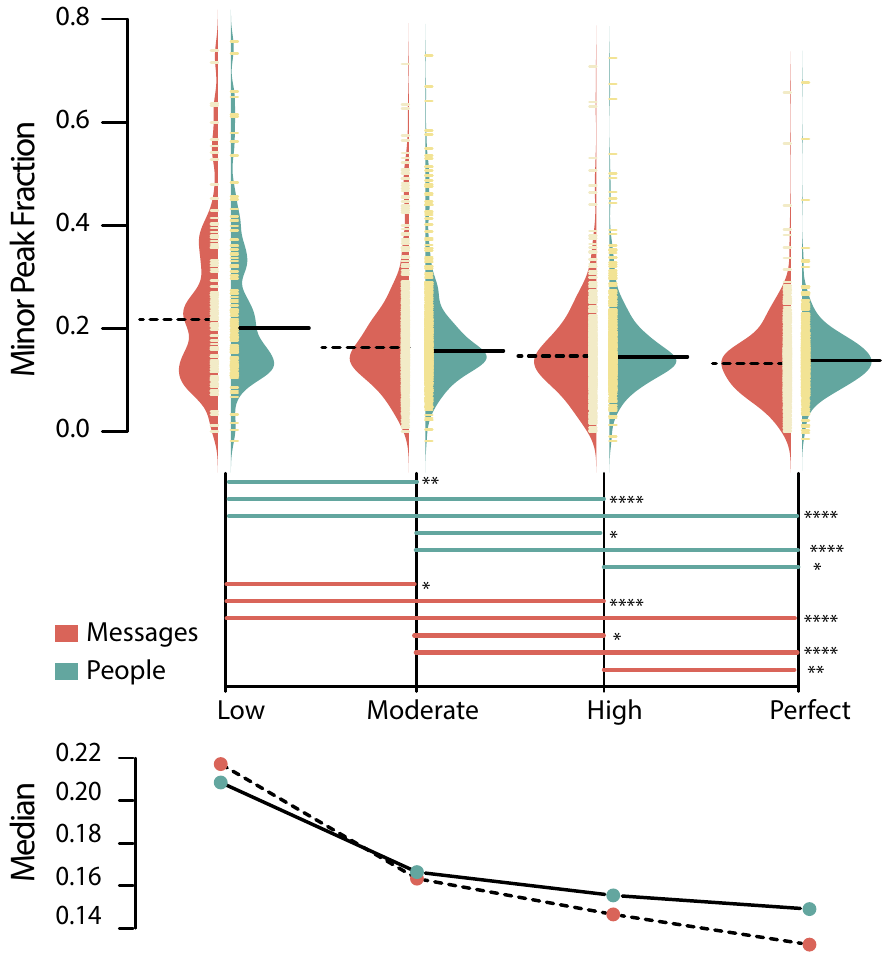}
 \label{fig:minorPeakFrac}
  \vspace{-10pt}
 \caption{}
\end{subfigure}
~
\begin{subfigure}[b]{0.49\textwidth}
 \includegraphics[width=\linewidth]{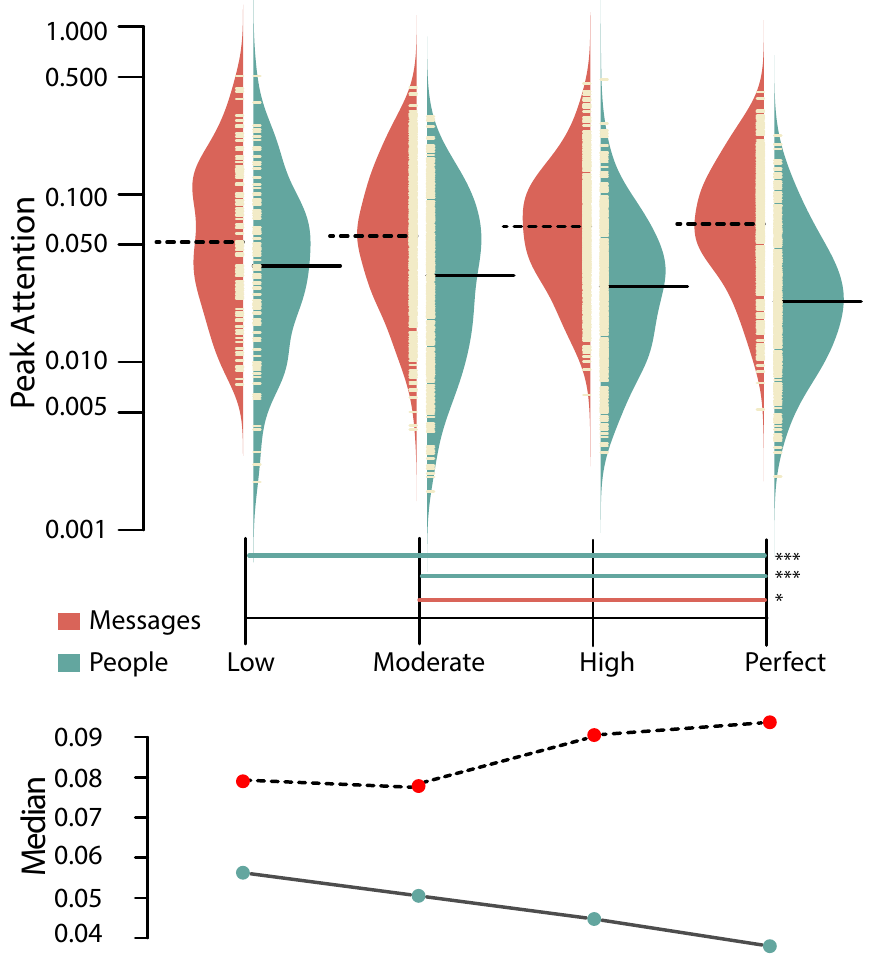}
 \label{fig:peakAttention}
  \vspace{-10pt}
 \caption{}
\end{subfigure}
 \vspace{-10pt}
\caption{Collective attention shown as a beanplot distribution. The shape of each half of the asymmetric bean represents the Gaussian density estimation of the distribution. The lines (in yellow) are the actual data points; the dotted long bean line is the median corresponding to the message volume, the solid line shows the median for people volume. The * denotes pairwise significant differences between cluster medians after correcting for familywise error-rate. (a).  Proportion of minor peak fractions are statistically different across all credibility class pairs for both message and unique people volume. (b). Peak attention is significantly different across ``Low'' and ``Perfect'', and ``Moderate'' and ``Perfect''  credibility classes for unique people volume, and ``Perfect'' and ``Moderate'' classes for message volume. The line charts at the bottom panel show the median trends across the credibility classes.}
\label{fig:peakyMsgs}
\end{figure*}

\subsection{Statistical Analysis and Results}
We tested the differences in collective attention measures across the credibility classes using  the Wilcoxon Rank Sum or Mann-Whitney U test. For each temporal measure (peak and minor peak attention) and for each collective attention metric (message and people volume), we performed pairwise Wilcoxon Rank Sum tests, followed by Bonferroni corrections \cite{dunn1959estimation} to control for potential inflation of the family-wise error rate by multiple test comparisons. 
We found that, for both \emph{message volume} and \emph{people volume}, differences in the minor peak fraction are statistically significant ($p < 0.00833$ after Bonferroni corrections and using Wilcoxon Rank Sum tests). As shown in Figure \ref{fig:peakyMsgs},  median \emph{minor peak attention} decreases as credibility level increases from ``Low'' to ``Perfect''. We also found a significant moderate degree of negative correlation between $P_{ca}$ and minor peak fraction for both \emph{message volume} ($r = -0.33$) and \emph{people volume} ($r = -0.33$).
These results suggest that an event attracting renewed interest is associated with lower perceived credibility. 
On the other hand, \emph{peak attention} of messages was only statistically different between ``Perfect'' and ``Moderate'' credibility classes. Peak attention for \emph{people volume} could only provide coarse-grained information separating ``Low'' and ``Perfect'', and ``Moderate'' and ``Perfect'' credibility classes. 
These results indicate that \emph{peak attention} is not a useful signal for event credibility.
To ensure that these ratio-based, collective attention measures described above are not sensitive to event duration, which can affect the denominator (cumulative volume), we performed pairwise Wilcoxon Rank Sum test comparisons of event duration across the credibility classes. We found no significant difference, indicating that event duration does not skew the collective attention metrics for a particular credibility class. Moreover, to ensure that our collective attention metrics are independent observations over time--a criteria necessary for the validity of our statistical analysis--we performed Ljung-Box Q (LBQ) tests \cite{ljung1978measure}. We were able to reject null hypothesis for each of our LBQ tests; thus confirming that our collective attention measures for both message and people volume are independent over time. 

\begin{figure*}[!t]
\centering
\begin{subfigure}[b]{0.49\textwidth}
 \includegraphics[scale=0.6]{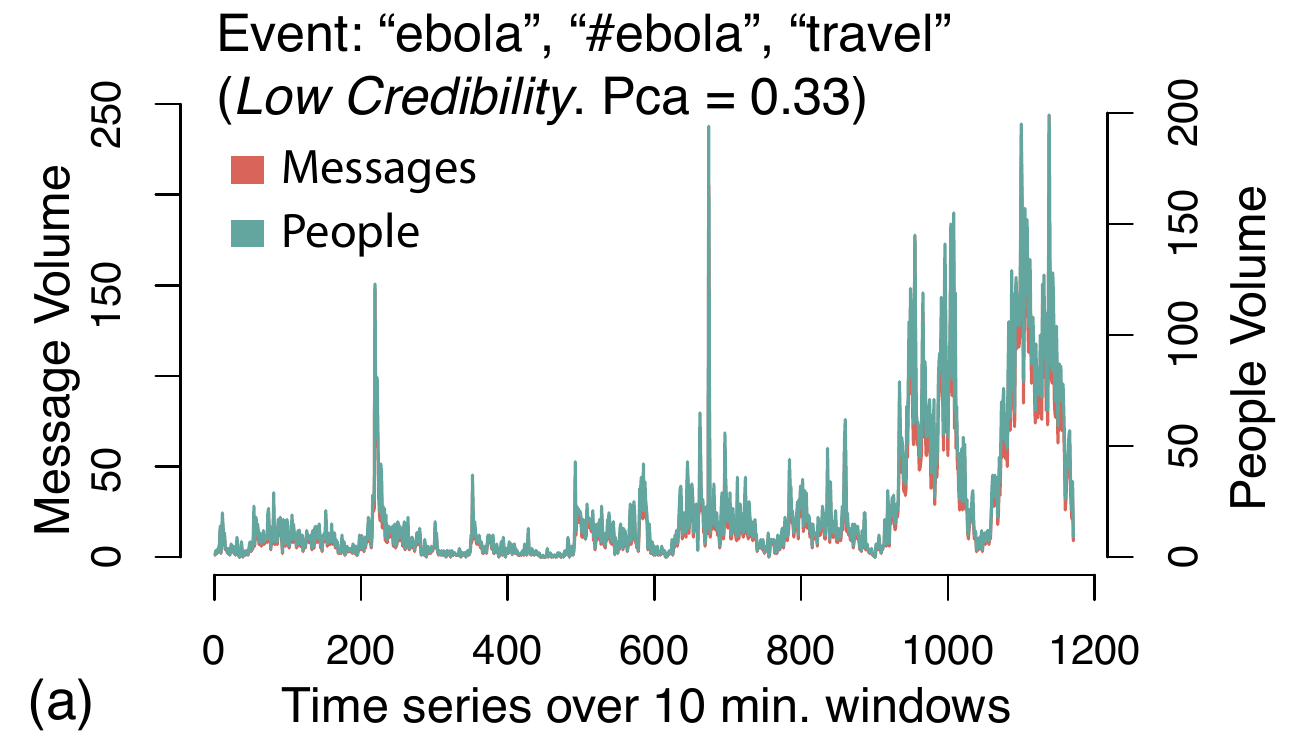}
 \label{fig:cred_low}
\end{subfigure}
~
\begin{subfigure}[b]{0.49\textwidth}
 \includegraphics[scale=0.6]{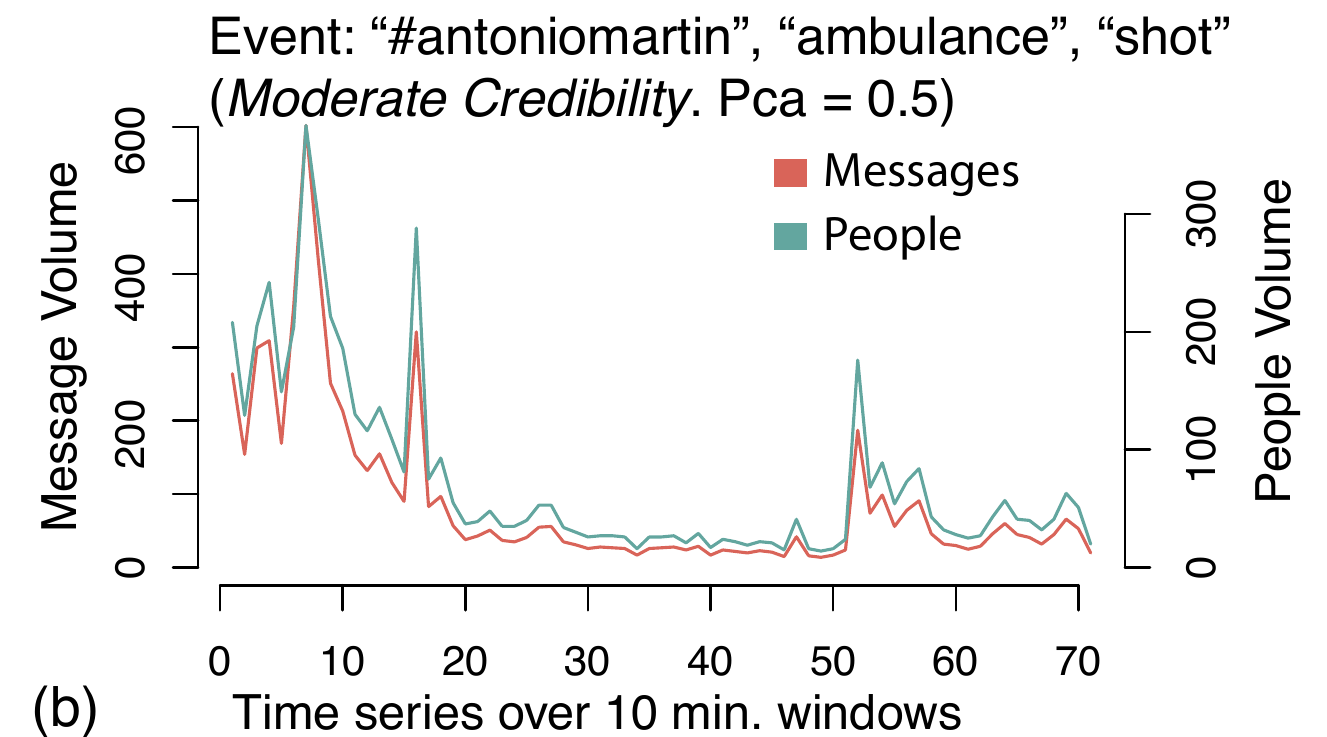}
 \label{fig:cred_mod}
\end{subfigure}

\begin{subfigure}[b]{0.49\textwidth}
 \includegraphics[scale=0.6]{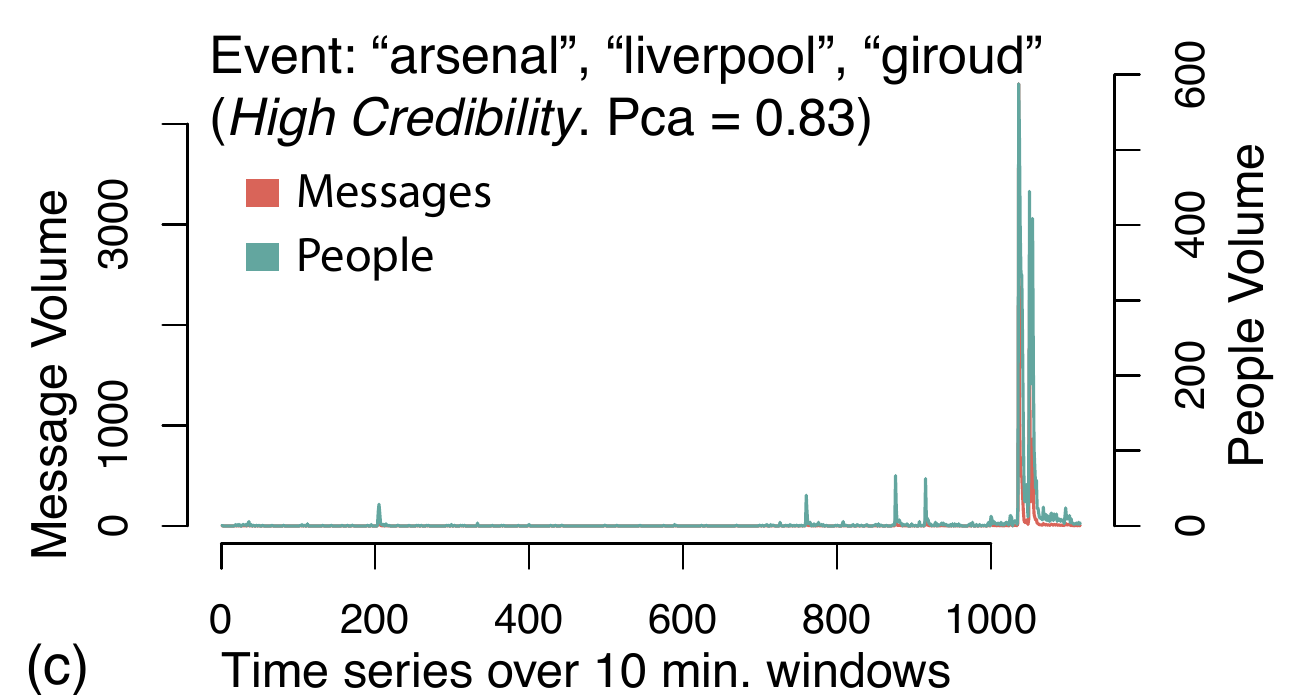}
 \label{fig:cred_high}
\end{subfigure}
~
\begin{subfigure}[b]{0.49\textwidth}
 \includegraphics[scale=0.6]{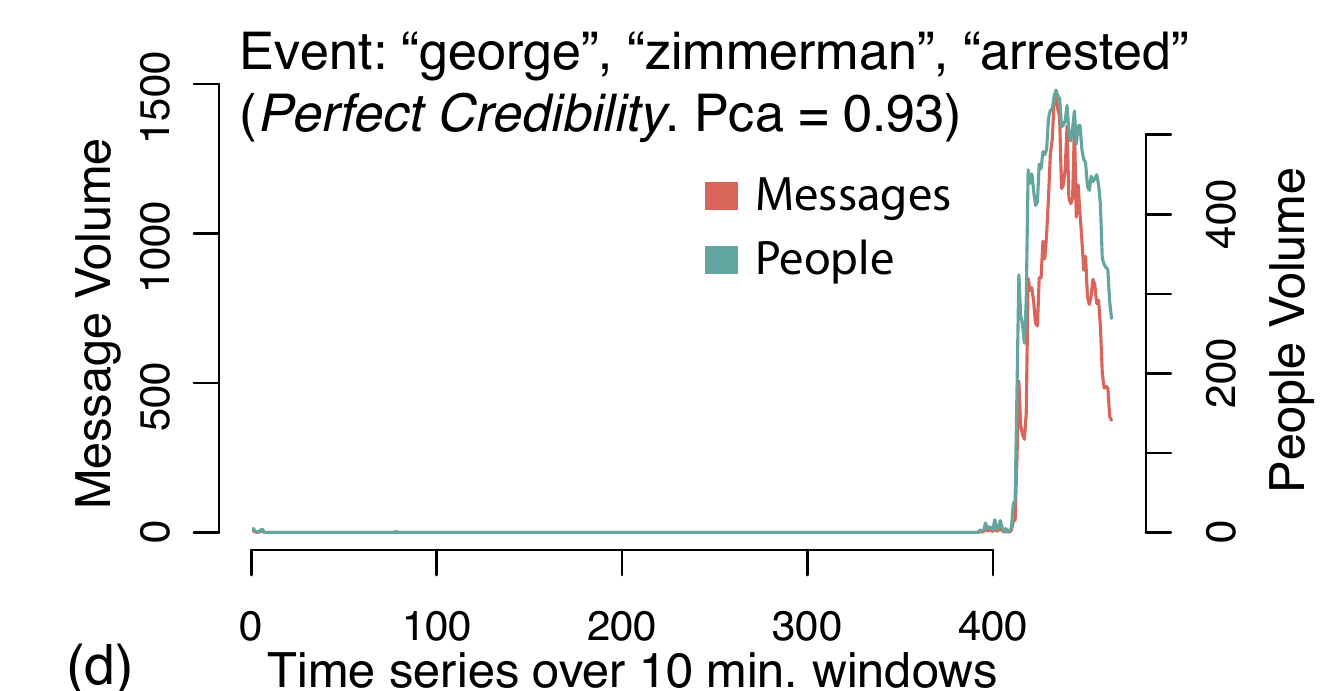}
 \label{fig:cred_perfect}
\end{subfigure}
\caption{Time series of collective attention metrics (message volume and unique people volume) for example events in each credibility class. The examples show representative behavior of collective attention metrics in each credibility class. While events in all four classes are marked by peak attention with respect to both message and people volume, events in the low and moderate credibility classes exhibit multiple minor peaks, signifying that persistent attention is characteristic of lower credible social media events.}
\label{fig:timeseries}
\vspace{-9pt}
\end{figure*}

\captionsetup[subfigure]{justification=justified}
\newcolumntype{C}[1]{>{\centering\let\newline\\\arraybackslash\hspace{0pt}}m{#1}}
\begin{table}
\small
\begin{subtable}{\textwidth}
\begin{tabular}{l | C{1.5cm}; 
{1pt/1.5pt} C{1.5 cm} ; 
{1pt/1.5pt} C{1.5 cm} ; 
{1pt/1.5pt} C{1.5 cm}} 
 & P & H & M & L \\ \hline
P & \cellcolor[HTML]{EFEFEF}{\color[HTML]{656565} } & {\color[HTML]{D96459} 0.00012**} & {\color[HTML]{D96459} 5.4e-12****} & {\color[HTML]{D96459} 4.6e-11****} \\ \hdashline
H & {\color[HTML]{63A69F} 0.002639*} & \cellcolor[HTML]{EFEFEF} & {\color[HTML]{D96459} 0.001703*} & {\color[HTML]{D96459} 3.6e-06****} \\ \hdashline
M & {\color[HTML]{63A69F} 1.0e-08****} & {\color[HTML]{63A69F} 0.004221**} & \cellcolor[HTML]{EFEFEF}{\color[HTML]{656565 } } & {\color[HTML]{D96459} 0.00547*} \\ \hdashline
L & {\color[HTML]{63A69F} 8.4e-11****} & {\color[HTML]{63A69F} 1.5e-06****} & {\color[HTML]{63A69F} 0.001981**} & \cellcolor[HTML]{EFEFEF}
\end{tabular}
\subcaption{Minor peak attention pairwise statistical differences}
\end{subtable}%

\vspace{8pt}
\begin{subtable}{\textwidth}
\begin{tabular}{l | C{1.5cm}; 
{1pt/1.5pt} C{1.5 cm} ; 
{1pt/1.5pt} C{1.5 cm} ; 
{1pt/1.5pt} C{1.5 cm}} 
 & P & H & M & L \\ \hline
P & \cellcolor[HTML]{EFEFEF}{\color[HTML]{656565} } & {\color[HTML]{D96459} ns} & {\color[HTML]{D96459}    0.00322*   } & {\color[HTML]{D96459}    ns   } \\ \hdashline
H & {\color[HTML]{63A69F}    ns   } & \cellcolor[HTML]{EFEFEF} & {\color[HTML]{D96459}    ns    } & {\color[HTML]{D96459}    ns   } \\ \hdashline
M & {\color[HTML]{63A69F} 0.00093***} & {\color[HTML]{63A69F}    ns   } & \cellcolor[HTML]{EFEFEF}{\color[HTML]{000000} } & {\color[HTML]{D96459} ns} \\ \hdashline
L & {\color[HTML]{63A69F} 0.00045***} & {\color[HTML]{63A69F} ns} & {\color[HTML]{63A69F}    ns   } & \cellcolor[HTML]{EFEFEF}
\end{tabular}
\subcaption{Peak attention pairwise statistical differences}
\end{subtable}
\caption{Pairwise statistical significance after Wilcoxon Rank Sum tests. P, H, M, L correspond to Perfect, High, Moderate and Low credibility classes. The top half of the diagonal corresponds to \emph{message volume}, while bottom half shows pairwise differences in \emph{people volume}. ns stands for non-significance.}
\vspace{-15pt}
\label{tab:wilcox}
\end{table}

\vspace{-5pt}
\section{Discussion}
By investigating the most comprehensive large-scale longitudinal credibility corpus constructed to date, we were able to test the relationship between an event's perceived credibility level and the temporal dynamics of its collective attention. According to our findings, moments of renewed collective attention are associated with event reportage marked by decreased levels of perceived credibility. 
Do frequent peaks in collective attention lead to lower perceived credibility? Or do reduced levels of credibility spark the continued interest in the event? Our current study cannot establish the causal direction of this relation. However, we are able to establish that the persistence of collective attention peaks is a reliable temporal signature for an event's perceived credibility level. 
Moreover, an advantage of viewing these phenomena through the lens of a fundamental property of human activity, such as time, is that the resultant findings are likely to hold irrespective of the platform (e.g., Twitter) hosting the collective human attention directed toward real-world events.

We remark that by using a simple proportion based classification technique we identified robust and scalable credibility classes; hence it is also potentially applicable to other online settings where user's collective attention drives popularity of content.  
Moreover, by using simple and interpretable parameters computed on times series of minute-wise user and message attention, we revealed vital temporal indicators associated with information credibility. Contrary to other sophisticated methods which require the estimation of power-law exponents for unraveling collective attention dynamics, or the calculation of costly correlations between activity time series, the parameters employed here can easily be computed in a scalable way. Although devoid of any predictive power, these measures can support the discovery of collective attention patterns in large-scale records of human activity. 


On the basis of these results, we envision that organizations struggling to handle the propagation of online misinformation \cite{time} can harness the temporality of collective attention to predict the level of credibility. We may be able to subsequently design interventions aimed at controlling the spread of false information or cautioning social media users to be skeptical about an evolving topic's veracity, ultimately raising an individual's capacity to assess credibility of information. 
Imagine a news reporting tool which shows social media discussions highlighting areas which witnessed multiple minor peaks of human activity, or think of a fact-checking system that compares temporal regions of high minor peak attentions to those with fewer attention peaks, or consider temporal tagging of scientific discourse or medical records emphasizing areas that garnered intermittent temporal popularity. We foresee that our findings can lead to a new class of such temporally aware systems which underscore degrees of information uncertainty based on temporal signals of collective attention. 
Finally, our study has practical implications in the field of computational social science where inferences about human social behavior are based on reports of online interactions \cite{counts2014computational} and trusting the credibility of those reports is crucial for any downstream analysis. For example, imagine a health researcher investigating the spread of Ebola via social media reports or a financial trader gauging market volatility based on breaking news and citizen reports on social media; veracity of those reports will affect the subsequent inferences. 

\section{Conclusions}
To study the dynamics of collective attention and its relation to information credibility in a natural setting, we analyzed the temporal patterns of 47M Twitter messages spread across 1,138 social media events along with their in-situ credibility ratings. We do so by multiple statistical comparison tests over parameters computed on the time series of collective attention of messages and distinct users. Although simple, this approach provides fundamental insights about collective attention and information credibility that would otherwise be missed by more complicated predictive analysis methods. 


%
\bibliographystyle{abbrv}
\balance
\bibliography{temporal-bib}  

\begin{thebibliography}{10}

\bibitem{sydney2016}
A.~Arif, K.~Shanahan, F.~Chou, Y.~Dosouto, K.~Starbird, and E.~S. Spiro.
\newblock How information snowballs: Exploring the role of exposure in online
  rumor propagation.
\newblock In {\em Proc. CSCW}, pages 465--476, 2016.

\bibitem{asur2011trends}
S.~Asur, B.~A. Huberman, G.~Szabo, and C.~Wang.
\newblock Trends in social media: Persistence and decay.
\newblock {\em Available at SSRN Scholarly Paper ID 1755748, Social Science
  Research Network.}, 2011.

\bibitem{barabasi2005origin}
A.-L. Barabasi.
\newblock The origin of bursts and heavy tails in human dynamics.
\newblock {\em Nature}, 435(7039):207--211, 2005.

\bibitem{bass1969frank}
M.~Bass.
\newblock Frank.
\newblock {\em A New Product Growth for Model Consumer Durables},
  50:1825--1832, 1969.

\bibitem{bordia1998rumor}
P.~Bordia and R.~L. Rosnow.
\newblock Rumor rest stops on the information highway transmission patterns in
  a computer-mediated rumor chain.
\newblock {\em Human Communication Research}, 25(2):163--179, 1998.

\bibitem{castillo2011information}
C.~Castillo, M.~Mendoza, and B.~Poblete.
\newblock Information credibility on twitter.
\newblock In {\em Proc. WWW}, 2011.

\bibitem{pew}
A.~Caumont.
\newblock \textit{12 trends shaping digital news} {(Pew Research Center)}.
\newblock
  \url{http://www.pewresearch.org/fact-tank/2013/10/16/12-trends-shaping-digital-news},
  2013.

\bibitem{chu2010tweeting}
Z.~Chu, S.~Gianvecchio, H.~Wang, and S.~Jajodia.
\newblock Who is tweeting on twitter: human, bot, or cyborg?
\newblock In {\em Proc. ACSAC}, pages 21--30. ACM, 2010.

\bibitem{counts2014computational}
S.~Counts, M.~De~Choudhury, J.~Diesner, E.~Gilbert, M.~Gonzalez, B.~Keegan,
  M.~Naaman, and H.~Wallach.
\newblock Computational social science: Cscw in the social media era.
\newblock In {\em Proc. CSCW Companion publication}, pages 105--108. ACM, 2014.

\bibitem{crane2008robust}
R.~Crane and D.~Sornette.
\newblock Robust dynamic classes revealed by measuring the response function of
  a social system.
\newblock {\em PNAS}, 105(41):15649--15653, 2008.

\bibitem{del2016spreading}
M.~Del~Vicario, A.~Bessi, F.~Zollo, F.~Petroni, A.~Scala, G.~Caldarelli, H.~E.
  Stanley, and W.~Quattrociocchi.
\newblock The spreading of misinformation online.
\newblock {\em PNAS}, 113(3):554--559, 2016.

\bibitem{dunn1959estimation}
O.~J. Dunn.
\newblock Estimation of the medians for dependent variables.
\newblock {\em The Annals of Mathematical Statistics}, pages 192--197, 1959.

\bibitem{ladaicwsm}
A.~Friggeri, L.~Adamic, D.~Eckles, and J.~Cheng.
\newblock Rumor cascades.
\newblock In {\em Proc. ICWSM}, 2014.

\bibitem{time_chappell}
S.~Frizell.
\newblock \textit{3 Muslim Students Murdered in North Carolina}.
\newblock \url{time.com/3704759/muslim-students-murdered-chapel-hill}, 2015.

\bibitem{goel2010predicting}
S.~Goel, J.~M. Hofman, S.~Lahaie, D.~M. Pennock, and D.~J. Watts.
\newblock Predicting consumer behavior with web search.
\newblock {\em PNAS}, 107(41):17486--17490, 2010.

\bibitem{golder2011diurnal}
S.~A. Golder and M.~W. Macy.
\newblock Diurnal and seasonal mood vary with work, sleep, and daylength across
  diverse cultures.
\newblock {\em Science}, 333(6051):1878--1881, 2011.

\bibitem{gruhl2004information}
D.~Gruhl, R.~Guha, D.~Liben-Nowell, and A.~Tomkins.
\newblock Information diffusion through blogspace.
\newblock In {\em Proc. WWW}, pages 491--501. ACM, 2004.

\bibitem{gupta2014tweetcred}
A.~Gupta, P.~Kumaraguru, C.~Castillo, and P.~Meier.
\newblock Tweetcred: Real-time credibility assessment of content on twitter.
\newblock In {\em International Conference on Social Informatics}, pages
  228--243. Springer, 2014.

\bibitem{haight1967handbook}
F.~A. Haight and F.~A. Haight.
\newblock Handbook of the poisson distribution.
\newblock 1967.

\bibitem{howell2013digital}
L.~Howell.
\newblock Digital wildfires in a hyperconnected world.
\newblock {\em World Economic Forum}, 2013.

\bibitem{hu2012breaking}
M.~Hu, S.~Liu, F.~Wei, Y.~Wu, J.~Stasko, and K.-L. Ma.
\newblock Breaking news on twitter.
\newblock In {\em Proc. CHI}, pages 2751--2754. ACM, 2012.

\bibitem{kwak2010twitter}
H.~Kwak, C.~Lee, H.~Park, and S.~Moon.
\newblock What is twitter, a social network or a news media?
\newblock In {\em Proc. WWW}, pages 591--600. ACM, 2010.

\bibitem{lehmann2012dynamical}
J.~Lehmann, B.~Gon{\c{c}}alves, J.~J. Ramasco, and C.~Cattuto.
\newblock Dynamical classes of collective attention in twitter.
\newblock In {\em Proc. WWW}, pages 251--260. ACM, 2012.

\bibitem{leskovec2007dynamics}
J.~Leskovec, L.~A. Adamic, and B.~A. Huberman.
\newblock The dynamics of viral marketing.
\newblock {\em ACM Transactions on the Web (TWEB)}, 1(1):5, 2007.

\bibitem{leskovec2009meme}
J.~Leskovec, L.~Backstrom, and J.~Kleinberg.
\newblock Meme-tracking and the dynamics of the news cycle.
\newblock In {\em Proc. SIGKDD}, pages 497--506. ACM, 2009.

\bibitem{liu2014rumors}
F.~Liu, A.~Burton-Jones, and D.~Xu.
\newblock Rumors on social media in disasters: Extending transmission to
  retransmission.
\newblock In {\em Proc. PACIS}, 2014.

\bibitem{ljung1978measure}
G.~M. Ljung and G.~E. Box.
\newblock On a measure of lack of fit in time series models.
\newblock {\em Biometrika}, 65(2):297--303, 1978.

\bibitem{time}
V.~Luckerson.
\newblock \textit{Fear, Misinformation, and Social Media Complicate Ebola
  Fight.}{ Time Inc.}
\newblock \url{time.com/3479254/ebola-social-media/}, 2014.

\bibitem{maddock2015characterizing}
J.~Maddock, K.~Starbird, H.~J. Al-Hassani, D.~E. Sandoval, M.~Orand, and R.~M.
  Mason.
\newblock Characterizing online rumoring behavior using multi-dimensional
  signatures.
\newblock In {\em Proc. CSCW}, pages 228--241. ACM, 2015.

\bibitem{maimon2005data}
O.~Maimon and L.~Rokach.
\newblock {\em Data mining and knowledge discovery handbook}, volume~2.
\newblock Springer, 2005.

\bibitem{mathioudakis2010twittermonitor}
M.~Mathioudakis and N.~Koudas.
\newblock Twittermonitor: trend detection over the twitter stream.
\newblock In {\em Proc. SIGMOD}, pages 1155--1158. ACM, 2010.

\bibitem{mitra2015credbank}
T.~Mitra and E.~Gilbert.
\newblock Credbank: A large-scale social media corpus with associated
  credibility annotations.
\newblock In {\em Proc. ICWSM}, 2015.

\bibitem{mitra2015comparing}
T.~Mitra, C.~J. Hutto, and E.~Gilbert.
\newblock Comparing person-and process-centric strategies for obtaining quality
  data on amazon mechanical turk.
\newblock In {\em Proc. CHI}, pages 1345--1354. ACM, 2015.

\bibitem{mitra2017credlang}
T.~Mitra, G.~Wright, and E.~Gilbert.
\newblock A parsimonious language model of social media credibility across
  disparate events.
\newblock In {\em Proc. CSCW}, 2017.

\bibitem{morris2012tweeting}
M.~R. Morris, S.~Counts, A.~Roseway, A.~Hoff, and J.~Schwarz.
\newblock Tweeting is believing?: understanding microblog credibility
  perceptions.
\newblock In {\em Proc. CSCW}, pages 441--450. ACM, 2012.

\bibitem{naaman2010really}
M.~Naaman, J.~Boase, and C.-H. Lai.
\newblock Is it really about me?: message content in social awareness streams.
\newblock In {\em Proc. CSCW}, pages 189--192. ACM, 2010.

\bibitem{o2010tweetmotif}
B.~O'Connor, M.~Krieger, and D.~Ahn.
\newblock Tweetmotif: Exploratory search and topic summarization for twitter.
\newblock In {\em Proc. ICWSM}, 2010.

\bibitem{qazvinian2011rumor}
V.~Qazvinian, E.~Rosengren, D.~R. Radev, and Q.~Mei.
\newblock Rumor has it: Identifying misinformation in microblogs.
\newblock In {\em Proc. EMNLP}, 2011.

\bibitem{ratkiewicz2010characterizing}
J.~Ratkiewicz, S.~Fortunato, A.~Flammini, F.~Menczer, and A.~Vespignani.
\newblock Characterizing and modeling the dynamics of online popularity.
\newblock {\em Physical review letters}, 105(15):158701, 2010.

\bibitem{romero2011differences}
D.~M. Romero, B.~Meeder, and J.~Kleinberg.
\newblock Differences in the mechanics of information diffusion across topics:
  idioms, political hashtags, and complex contagion on twitter.
\newblock In {\em Proc. WWW}, pages 695--704. ACM, 2011.

\bibitem{salganik2006experimental}
M.~J. Salganik, P.~S. Dodds, and D.~J. Watts.
\newblock Experimental study of inequality and unpredictability in an
  artificial cultural market.
\newblock {\em Science}, 311(5762):854--856, 2006.

\bibitem{sasahara2013quantifying}
K.~Sasahara, Y.~Hirata, M.~Toyoda, M.~Kitsuregawa, and K.~Aihara.
\newblock Quantifying collective attention from tweet stream.
\newblock {\em PloS one}, 8(4):e61823, 2013.

\bibitem{shibutani1966improvised}
T.~Shibutani.
\newblock {\em Improvised news: A sociological study of rumor}.
\newblock Ardent Media, 1966.

\bibitem{sinha2010econophysics}
S.~Sinha, A.~Chatterjee, A.~Chakraborti, and B.~K. Chakrabarti.
\newblock {\em Econophysics: an introduction}.
\newblock John Wiley \& Sons, 2010.

\bibitem{stewart2015calculus}
J.~Stewart.
\newblock {\em Calculus: {E}arly transcendentals}.
\newblock Cengage Learning, 2015.

\bibitem{tibshirani2005cluster}
R.~Tibshirani and G.~Walther.
\newblock Cluster validation by prediction strength.
\newblock {\em Journal of Computational and Graphical Statistics},
  14(3):511--528, 2005.

\bibitem{tufekci2014big}
Z.~Tufekci.
\newblock Big questions for social media big data: Representativeness, validity
  and other methodological pitfalls.
\newblock In {\em In Proc. ICWSM}, 2014.

\bibitem{wang2007mining}
X.~Wang, C.~Zhai, X.~Hu, and R.~Sproat.
\newblock Mining correlated bursty topic patterns from coordinated text
  streams.
\newblock In {\em Proc. SIGKDD}, pages 784--793. ACM, 2007.

\bibitem{ward1963hierarchical}
J.~H. Ward~Jr.
\newblock Hierarchical grouping to optimize an objective function.
\newblock {\em Journal of the American statistical association},
  58(301):236--244, 1963.

\bibitem{wu2007novelty}
F.~Wu and B.~A. Huberman.
\newblock Novelty and collective attention.
\newblock {\em PNAS}, 104(45):17599--17601, 2007.

\bibitem{wu2010evidence}
Y.~Wu, C.~Zhou, J.~Xiao, J.~Kurths, and H.~J. Schellnhuber.
\newblock Evidence for a bimodal distribution in human communication.
\newblock {\em PNAS}, 107(44):18803--18808, 2010.

\bibitem{yang2011patterns}
J.~Yang and J.~Leskovec.
\newblock Patterns of temporal variation in online media.
\newblock In {\em Proc. WWW}, pages 177--186. ACM, 2011.

\bibitem{zeng2016unconfirmed}
L.~Zeng, K.~Starbird, and E.~S. Spiro.
\newblock \# unconfirmed: Classifying rumor stance in crisis-related social
  media messages.
\newblock In {\em Proc. ICWSM}, 2016.

\end{thebibliography}

%
%
\vspace{-5pt}
\appendix
\section{CREDBANK Construction Steps}
\noindent CREDBANK was constructed by following a sequence of phases:
\noindent
\textbf{1. Streaming Tweets and Preprocessing}: Twitter's Streaming API was used to iteratively collect a continuous 1\% sample of all global tweets, filtered to contain only English tweets, followed by spam removal, tokenization using a Twitter specific tokenizer \cite{o2010tweetmotif} and a sophisticated multi-stage stop word removal step.

\noindent
\textbf{2. Detecting Event Candidates}: After carefully considering various approaches for event detection from social media streams, we opted for topic models, since topic models can learn term co-occurences and unlike keyword based techniques do not make a-priori assumptions about what constitutes an event.

\noindent
\textbf{3. Event annotation}: To eliminate detection of potential false positives using a purely computational event detection approach, candidate events from the previous step were sent to ten independent human raters from Amazon Mechanical Turk (AMT) for judging whether a topic relates to a real-world news event. The majority agreement was selected as the final annotation.

\noindent
\textbf{4. Credibility Assessment}: This phase had three primary steps: 
\begin{itemize} [leftmargin=*,noitemsep,nolistsep, topsep=0pt, parsep=0pt,partopsep=0pt]
\itemsep0em 
\item[] \emph{Determining the credibility scale}:  Informed by work done by the linguistic community on `Event Factuality', the credibility scale was designed as an interaction between two dimensions: Polarity, which differentiates among `Accurate', `Inaccurate', and `Uncertain', and Degree of certainty which distinguishes among `Certainly',`Probably' and `Uncertain', leading to a 5-point Likert scale annotation scheme.
\item[] \emph{Determining number of independent Turk ratings for high quality annotation}:  
In this step, we piloted the CREDBANK system for 5 days collecting and annotating 50 events by both Turkers and expert annotators (university research librarians). The pilot study was followed by computing correlation statistics between Turker mean responses and expert mean responses while varying the count of independent Turker ratings per event. The correlation maximized at 30 Turker ratings leading to the decision of collecting 30 annotations per event.
\item[] \emph{Credibility assessment task}: The credibility assessment task framework was designed to ensure that the collected credibility ratings is of high quality. Multiple controlled experiments were performed before finalizing the strategy best suited for obtaining quality annotations \cite{mitra2015comparing}. Turkers were first selectively screened and trained via a qualification test. Screened workers were then directed to a task interface and asked to categorize an event's credibility after reading through a stream of real-time tweets related to an event topic. They were instructed to either be knowledgeable on the event topic or search online before making their credibility judgments. 

\textbf{5. Collecting Event Streams}: The final phase used Twitter's search API to collect all tweets specific to the event topic.
\end{itemize}

\end{document}